\documentclass[11pt]{article}
\usepackage[margin=1in]{geometry}
\usepackage{amsmath,amssymb,amsthm}
\usepackage{graphicx}
\usepackage{booktabs}
\usepackage{caption}
\usepackage{subcaption}
\usepackage{xcolor}
\usepackage[colorlinks=true,linkcolor=blue,citecolor=blue,urlcolor=blue]{hyperref}
\usepackage{natbib}
\usepackage[protrusion=true,expansion=false]{microtype}
\usepackage{multirow}

\newtheorem{conjecture}{Conjecture}

\newcommand{\NE}{\mathcal{N}}
\newcommand{\KL}{\mathrm{KL}}
\newcommand{\HH}{\mathcal{H}}

\newcommand{\argmin}{\operatorname*{arg\,min}}
\newcommand{\argmax}{\operatorname*{arg\,max}}
\newcommand{\Iproj}{\Pi}

\title{\bf Which Nash Equilibrium? \\ Solver-Dependent Selection on Zero-Sum Nash Polytopes}
\author{Luis Leal\\ \texttt{wichofer89@gmail.com}}
\date{}

\begin{document}
\maketitle

\begin{abstract}
\noindent
Many two-player zero-sum games admit not a unique Nash equilibrium but a convex \emph{set} of them: a polytope of profiles that all share the game's minimax value $V^\star$ yet prescribe materially different behaviour. Standard solvers are each guaranteed to converge to \emph{some} equilibrium and are commonly treated as interchangeable. We ask whether they instead select \emph{different} members of the Nash set, systematically as a function of the algorithm rather than the random seed. Using a tabular, exactly solvable testbed of six games with analytically known Nash sets---including a game with a two-dimensional Nash polytope and Kuhn poker---we find that (i)~equilibrium selection is determined by the algorithm, not the seed, but the differences between algorithm families emerge only on geometrically \emph{asymmetric} Nash sets; (ii)~regularized last-iterate methods (R-NaD, magnetic mirror descent) select the \emph{maximum-entropy} member, coinciding with the information projection of their uniform reference onto the Nash set---exactly on the 2-D polytope and at $99.7\%$ of maximum entropy in Kuhn---while regret-averaging methods (CFR, CFR\textsuperscript{+}, and fictitious play) drift to a lower-entropy face on the geometrically \emph{asymmetric} games---with fictitious play additionally drifting on the symmetric games and coinciding with max-entropy only in Kuhn, an unexplained single-game anomaly---whereas on the symmetric controls CFR and CFR\textsuperscript{+} agree with the max-entropy member; we confirm this split on a randomized $180$-game ensemble of asymmetric games, where R-NaD attains the maximum-entropy member in $100\%$ of converged games while CFR\textsuperscript{+} sits strictly below it in $94\%$ (paired Wilcoxon $p<10^{-27}$); (iii)~the selected member has downstream consequences against sub-optimal opponents that scale with sequential/hidden-information structure but remain bounded in magnitude---in Kuhn the max-entropy member is a strictly better hedge against flawed opponents, whereas on the matrix games the members differ without either dominating. We additionally report two negative results that correct common intuitions: removing CFR's positive-orthant ($\max(R,0)$) projection does \emph{not} eliminate boundary drift, so that projection is not its cause; and R-NaD's selection is \emph{anchor-following}, not initialization-independent. We state the maximum-entropy / I-projection characterization as a strongly data-supported \emph{conjecture} and delineate the open theory. All claims are checked against analytic ground truth.
\end{abstract}

\section{Introduction}
\label{sec:intro}

A central simplification in computational game theory is the treatment of a Nash equilibrium as a unique target: one solves a game, obtains \emph{the} equilibrium, and deploys it. For two-player zero-sum (2p0s) games this is justified at the level of \emph{value}---the minimax theorem guarantees a unique game value $V^\star$---but it is frequently false at the level of \emph{behaviour}. Degenerate or structurally symmetric games, and many games of practical interest, possess an entire convex set of Nash equilibria. Every member achieves $V^\star$ against a best-responding opponent, yet members can differ arbitrarily in how they behave off the equilibrium path, in the support they use, and in their robustness to imperfect opponents.

When the equilibrium set is not a singleton, a solver does not merely \emph{find} an equilibrium; it \emph{selects} one. Modern solvers span two broad families with very different update structure: \emph{regret-averaging} methods such as counterfactual regret minimization (CFR) and CFR\textsuperscript{+} \citep{zinkevich2007,tammelin2014,bowling2015}, whose deployed strategy is a time-average, and \emph{regularized last-iterate} methods such as Regularized Nash Dynamics (R-NaD), the engine behind DeepNash \citep{perolat2021,perolat2022}, and Magnetic Mirror Descent (MMD) \citep{sokota2023}, whose deployed strategy is the final iterate of a dynamics anchored to a reference policy. These families are routinely used interchangeably. We ask a simple question with a non-trivial answer:

\begin{quote}
\emph{When a 2p0s game has a set of Nash equilibria, do different solvers select different members of that set, and is the choice a systematic property of the algorithm?}
\end{quote}

We study this in a deliberately minimal, fully transparent setting: a tabular extensive-form engine using exact counterfactual values (no sampling or function-approximation variance), applied to six games whose Nash sets we know analytically. This lets us check every claim against ground truth rather than against another solver's output.

\paragraph{Contributions.}
\begin{enumerate}
\itemsep0.15em
\item \textbf{Selection is algorithmic and geometry-gated} (\S\ref{sec:core},\S\ref{sec:seed}). Across-algorithm differences are large on \emph{asymmetric} Nash sets and vanish on symmetric ones; the tabular solvers are deterministic, so within-algorithm variation across seeds is exactly zero, and the selection is invariant to the iteration budget (regret-averaging drift \emph{grows} with budget rather than closing, \S\ref{sec:seed}).
\item \textbf{A maximum-entropy / I-projection characterization for regularized methods} (\S\ref{sec:core}--\ref{sec:anchor}). With a uniform reference, R-NaD selects the maximum-entropy equilibrium---exactly on a 2-D polytope and at $99.7\%$ of maximum entropy in Kuhn---which equals the information projection of the reference onto the Nash set. We confirm this at population scale on a randomized $180$-game ensemble (\S\ref{sec:ensemble}): R-NaD attains the analytic max-entropy member in $100\%$ of converged games (median coordinate error $2\times10^{-4}$), with two independent degrees of freedom satisfied on the higher-dimensional faces, while CFR\textsuperscript{+} sits strictly below it (mean entropy gap $+0.121$, $95\%$ CI $[+0.10,+0.14]$). We show (\S\ref{sec:anchor}) this is \emph{anchor-following}: the selection moves with the initial reference, so the property holds \emph{under uniform initialization} rather than unconditionally. We state the characterization as a conjecture (\S\ref{sec:disc}).
\item \textbf{A controlled refutation of a common mechanism} (\S\ref{sec:hedge}). The boundary drift of CFR-family methods is widely attributed to the positive-orthant projection $\max(R,0)$ of regret matching. We test this by replacing the projection with a softmax (Hedge) while holding the regrets fixed; boundary drift \emph{increases}. The projection is therefore not the cause.
\item \textbf{Downstream consequences scale with structure but are bounded} (\S\ref{sec:teeth}). The selected member's robustness to flawed opponents differs by a factor of ${\sim}5.6\times$ between Kuhn and matrix games, but the absolute effect is small ($<0.02$). In Kuhn the max-entropy member is a strictly better hedge; on the matrix games the members differ without either dominating, so the hedging advantage tracks sequential structure rather than max-entropy itself. We make precise why genuinely payoff-inequivalent equilibria cannot arise in this setting.
\end{enumerate}

\noindent We emphasise throughout the distinction between \emph{phenomena we demonstrate} and \emph{mechanisms we conjecture}; two textbook mechanisms are refuted or downgraded by our experiments.

\section{Preliminaries}
\label{sec:prelim}

\paragraph{Games and equilibria.}
We consider finite two-player zero-sum games in extensive form with perfect recall; normal-form (matrix) games are the single-decision special case. Player $i$ has a behavioural strategy $\sigma_i$ assigning a distribution over actions to each of its information sets, and $\sigma=(\sigma_0,\sigma_1)$ denotes a profile. Let $u(\sigma)=u_0(\sigma)=-u_1(\sigma)$ be player $0$'s expected utility. A profile $\sigma^\star$ is a Nash equilibrium if neither player can improve by unilateral deviation. We write $\NE(G)$ for the set of Nash equilibria of $G$. By the minimax theorem, all $\sigma\in\NE(G)$ share the same value $u(\sigma)=V^\star$; moreover $\NE(G)$ is convex in sequence-form (realization) coordinates. We refer to $\NE(G)$ as the \emph{Nash polytope}.

\paragraph{Exploitability.}
Convergence is measured by exploitability (NashConv),
\begin{equation}
\textsc{NashConv}(\sigma)\;=\;\Big[\max_{\sigma_0'} u(\sigma_0',\sigma_1)-V^\star\Big]\;+\;\Big[V^\star-\min_{\sigma_1'} u(\sigma_0,\sigma_1')\Big],
\label{eq:nashconv}
\end{equation}
which is non-negative and zero exactly at Nash. We compute each best response exactly by enumeration over the (small) pure-strategy space, so \eqref{eq:nashconv} is exact.

\paragraph{The maximum-entropy member and the I-projection.}
Among all equilibria we single out the maximum-entropy member
\begin{equation}
\sigma^{\mathrm{ME}}\;=\;\argmax_{\sigma\in\NE(G)}\ \HH(\sigma),
\qquad \HH(\sigma)=\frac{1}{|\mathcal{I}|}\sum_{I\in\mathcal{I}} H\!\big(\sigma(\cdot\mid I)\big),
\label{eq:me}
\end{equation}
where $\mathcal I$ is the set of information sets and $H$ is Shannon entropy. Because $H(p)=-\KL(p\,\Vert\,\mathrm{unif})+\log|A|$, maximizing entropy over a convex set is equivalent to minimizing relative entropy to the uniform distribution: $\sigma^{\mathrm{ME}}$ is the \emph{information projection} (I-projection) of the uniform policy onto $\NE(G)$. More generally, for a reference $\rho$,
\begin{equation}
\Iproj_{\NE}(\rho)\;=\;\argmin_{\sigma\in\NE(G)}\ \KL\!\big(\sigma\,\Vert\,\rho\big),
\label{eq:iproj}
\end{equation}
with $\Iproj_{\NE}(\mathrm{unif})=\sigma^{\mathrm{ME}}$. The central empirical claim of this paper concerns whether regularized solvers compute \eqref{eq:iproj}. That entropic regularization selects the maximum-entropy / I-projection member of an optimal set is itself a well-established principle outside game solving: the vanishing-penalty limit of an entropically regularized linear program is the maximum-entropy point of the optimal face \citep{weed2018}, and entropic optimal transport analogously selects the KL projection of a reference onto the feasible polytope \citep{cuturi2013}. The maximum-entropy Nash equilibrium has likewise been singled out before as a canonical member, e.g.\ for its invariance to duplicated strategies \citep{balduzzi2018}. Our question is therefore not whether such a selection principle exists, but whether---and which of---the game solvers in standard use realize it.

\paragraph{Solvers.}
We use a single tabular engine with exact counterfactual values $q_t(I,\cdot)$ and reach probabilities, instantiating six update rules. Let $s=\sigma_t(\cdot\mid I)$ and $q=q_t(I,\cdot)$ at information set $I$.
\begin{itemize}
\itemsep0.2em
\item \textbf{CFR} \citep{zinkevich2007}: cumulative regret $R\!\mathrel{+}=\!q-s^\top q$; current policy $\sigma_{t+1}\propto\max(R,0)$; deploy the \emph{uniform} time-average.
\item \textbf{CFR\textsuperscript{+}} \citep{tammelin2014}: regret-matching\textsuperscript{+}, $R\leftarrow\max(R+(q-s^\top q),0)$; deploy the \emph{linearly weighted} average.
\item \textbf{Hedge} (used only as an ablation, \S\ref{sec:hedge}): identical counterfactual regrets but $\sigma_{t+1}\propto\exp(\eta R)$---a softmax with no positive-orthant projection; deploy the average.
\item \textbf{Fictitious play (FP)} \citep{brown1951,robinson1951}: realization-weighted averaging of exact best responses; deploy the average.
\item \textbf{Magnetic Mirror Descent (MMD)} \citep{sokota2023}: with reference $\rho$ and temperature $\lambda$, $c=1/(1+\eta\lambda)$,
\begin{equation}
\log\sigma_{t+1}\ \propto\ c\,\log s + c\,\eta\,q + (1-c)\,\log\rho .
\label{eq:mmd}
\end{equation}
We run MMD with a \emph{fixed} uniform reference (a small fixed magnet).
\item \textbf{R-NaD} \citep{perolat2021,perolat2022}: the same regularized update \eqref{eq:mmd} but with a \emph{moving} reference---$\rho$ is periodically reset to the current policy. The fixed-reference update converges to the quantal-response equilibrium (QRE) \citep{mckelvey1995} that is the regularized fixed point; resetting the reference traces a sequence of QREs whose limit is an unregularized Nash equilibrium. This moving-reference annealing is the algorithmic analogue of the logit-QRE homotopy, traced from the uniform distribution at infinite temperature, whose zero-temperature endpoint is the \emph{limiting logit equilibrium} \citep{mckelvey1995,turocy2005} and which underlies homotopy solvers such as ADIDAS \citep{gemp2022}. Deploy the final iterate.
\item \textbf{MWU / NeuRD} \citep{hennes2020}: the unregularized last iterate $\sigma_{t+1}\propto\exp(\log s+\eta q)$; deploy the final iterate.
\end{itemize}
The two families differ in what is deployed: regret-averaging methods (CFR, CFR\textsuperscript{+}, Hedge, FP) deploy a \emph{time-average}; regularized/unregularized last-iterate methods (MMD, R-NaD, MWU) deploy the \emph{final iterate}.

\section{Experimental setup}
\label{sec:setup}

\paragraph{Testbed.}
We use six games with analytically characterised Nash sets (exact payoffs are given in the released engine \texttt{efg.py}). Each game's Nash set is parameterised by a scalar \emph{selection coordinate} $c\in[0,1]$ (a normalisation of the relevant family parameter), except \texttt{polytope4} whose Nash set is genuinely two-dimensional.
\begin{itemize}
\itemsep0.2em
\item \texttt{pennies\_safe}, \texttt{two\_safe}: matching-pennies cores augmented with one or two value-preserving ``safe'' actions, giving a \emph{symmetric} 1-D Nash segment (max-entropy coordinate $0.333$ and $0.500$).
\item \texttt{dup\_action}: a duplicated-action degeneracy producing a 1-D segment (max-entropy $0.250$).
\item \texttt{asym\_safe}: an \emph{asymmetric} safe-action game whose Nash face is $P_0=(p_0,\,2p_0,\,1-3p_0)$, $p_0\in[0,\tfrac13]$ (max-entropy coordinate $0.218$).
\item \texttt{polytope4}: a game with a genuinely \emph{two-dimensional} Nash polytope (max-entropy point $(0.161,0.256)$).
\item \texttt{kuhn}: Kuhn poker, an imperfect-information extensive-form game with a one-parameter family of equilibria indexed by the first player's bluff frequency (max-entropy bluff $0.201$; game value $V^\star=-1/18$).
\end{itemize}
The symmetric games are controls (selection should be unambiguous); the asymmetric games and Kuhn are where families can disagree. Beyond these six fixed games, \S\ref{sec:ensemble} introduces a randomized ensemble of $180$ asymmetric matrix games---generated with analytically known Nash faces---to test whether the findings generalize beyond hand-built instances.

\paragraph{Metrics.}
For each (game, solver) we report the selection coordinate, the mean policy entropy \eqref{eq:me}, exploitability \eqref{eq:nashconv}, the average $L_0$ support (number of actions with probability $>10^{-4}$), and the mean Jensen--Shannon divergence of the profile to the R-NaD profile (a game-agnostic, coordinate-free measure of how far two solvers' selections lie apart). A solver is deemed \emph{converged} if its exploitability is below $0.02$.

\paragraph{Protocol.}
All solvers start from the uniform policy unless stated otherwise. Iteration budgets are fixed per solver and are not tuned per game. We emphasise that this fixed budget is not the reason the fixed-magnet MMD and MWU baselines fail to converge on the asymmetric games (\S\ref{sec:bakeoff}): their non-convergence is the expected limit-cycling of \emph{insufficiently regularized} (MMD, whose fixed magnet is too weak to stabilise the dynamics) and \emph{unregularized} (MWU) learning in zero-sum games, not an artifact of an under-resourced sweep---larger budgets do not remove it. We solve every (game, solver) pair once and reuse the solutions across all analyses. Ground truth is verified first (\S\ref{sec:groundtruth}); only then are selection inferences drawn.

\section{Results}

\subsection{Ground truth}
\label{sec:groundtruth}
Before asking which member each solver selects, we confirm the analytic Nash sets are genuinely Nash. Sweeping the Kuhn family over its parameter range yields a maximum exploitability of $<10^{-6}$, and the analytic max-entropy coordinates of \texttt{asym\_safe} ($0.218$) and \texttt{polytope4} ($(0.161,0.256)$) satisfy the equilibrium conditions exactly. The parameterised families moreover \emph{exhaust} the Nash set rather than tracing a sub-arc of it: in each matrix game player~1's equilibrium strategy is unique and player~0's indifference pins the active-row ratio, leaving only the safe-mass (or bluff) parameter free, so the max-entropy coordinate we report is the maximum over the \emph{entire} Nash set; numerically, every family member has exploitability ${<}10^{-15}$ while profiles that break the defining ratio constraint are exploitable (e.g.\ $0.08$ on \texttt{polytope4}). All subsequent comparisons are therefore against verified ground truth.

\subsection{Core result: selection across games}
\label{sec:core}

\begin{figure}[t]
\centering
\includegraphics[width=0.95\textwidth]{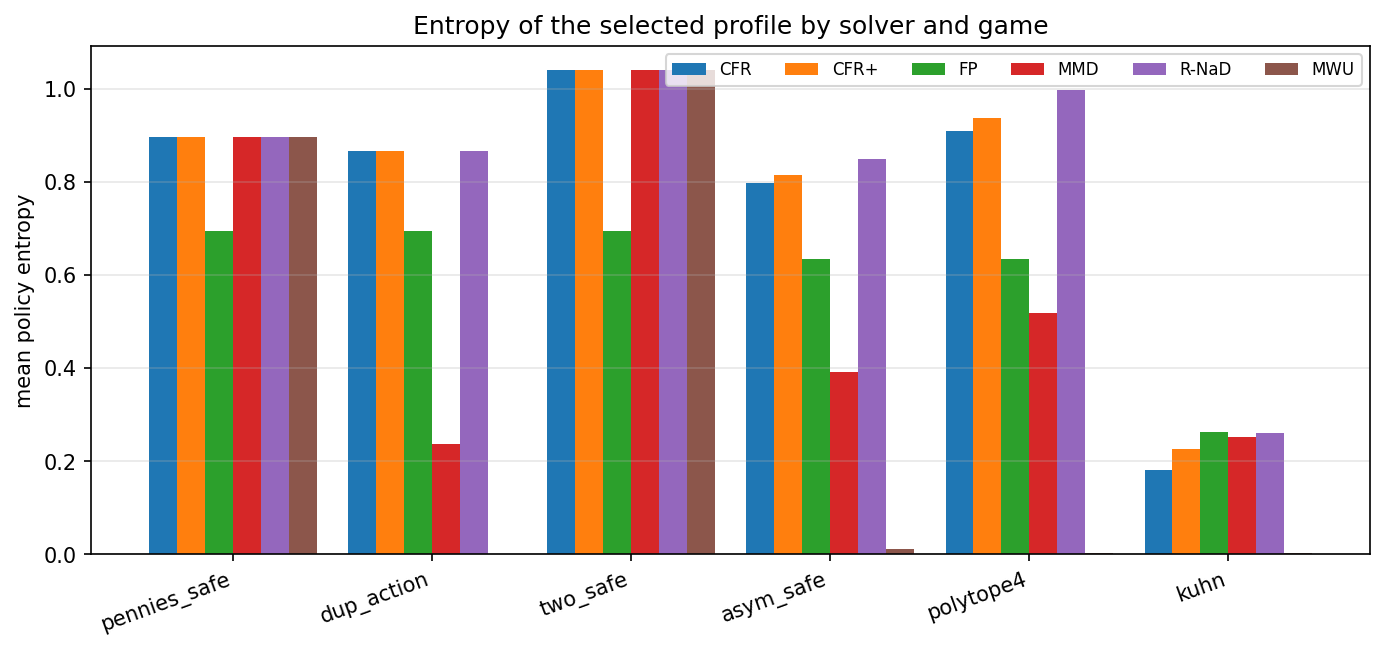}
\caption{Mean policy entropy of the selected profile, by solver and game. On symmetric games all converging solvers coincide; on the asymmetric games (\texttt{asym\_safe}, \texttt{polytope4}, \texttt{kuhn}) R-NaD attains the highest entropy while CFR/CFR\textsuperscript{+} sit lower and MWU collapses.}
\label{fig:entropy}
\end{figure}

Table~\ref{tab:core} reports the selection coordinate of every solver on every game; Figure~\ref{fig:entropy} shows the corresponding policy entropies. The structure is consistent and sharp.

On the \emph{symmetric} games (\texttt{pennies\_safe}, \texttt{two\_safe}, \texttt{dup\_action}) every converging solver \emph{except fictitious play} agrees on the max-entropy coordinate; FP is the lone exception, drifting to the boundary even here (e.g.\ $0.000$ on \texttt{pennies\_safe}, where the others select $0.333$), consistent with its boundary drift across all matrix games discussed in \S\ref{sec:disc}. On the \emph{asymmetric} games (\texttt{asym\_safe}, \texttt{polytope4}, \texttt{kuhn}) the families separate: R-NaD sits at (or, in Kuhn, very near) the max-entropy coordinate, while CFR and CFR\textsuperscript{+} drift to a distinctly lower-entropy interior/boundary point. On \texttt{asym\_safe}, R-NaD selects $0.218$ (exactly max-entropy) versus CFR\textsuperscript{+}'s $0.275$; on \texttt{polytope4}, R-NaD selects $0.162$ against the analytic $0.161$ while CFR\textsuperscript{+} selects $0.246$; on Kuhn, R-NaD selects bluff $0.180$ (entropy $0.261$ versus the max-entropy $0.262$) while CFR\textsuperscript{+} selects $0.068$. The unregularized last iterate (MWU) and the fixed-magnet MMD fail to converge on the asymmetric matrix games (\S\ref{sec:bakeoff}); their reported coordinates are artefacts of non-convergence and are flagged accordingly.

\begin{table}[t]
\centering
\caption{Selection coordinate by solver and game. ME $=$ analytic max-entropy coordinate. Bold marks agreement with ME to within $0.02$. A dagger ($\dagger$) marks solvers that did not reach exploitability $<0.02$ within the fixed budget (their coordinates reflect non-convergence, not selection). \texttt{polytope4} coordinates are player~0's probability on row action $r_0$ (the first coordinate $p_0$ of its 2-D selection). Mean policy entropy is shown in Figure~\ref{fig:entropy}; the full per-cell coordinate, entropy, and exploitability are in Appendix Table~\ref{tab:full}.}
\label{tab:core}
\small
\begin{tabular}{lccccccc}
\toprule
Game & ME & CFR & CFR\textsuperscript{+} & FP & MMD & R-NaD & MWU \\
\midrule
\texttt{pennies\_safe} & 0.333 & \textbf{0.333} & \textbf{0.333} & 0.000 & \textbf{0.333} & \textbf{0.333} & \textbf{0.333} \\
\texttt{dup\_action}   & 0.250 & \textbf{0.248} & \textbf{0.250} & 0.504 & 0.008$^\dagger$ & \textbf{0.250} & 0.000$^\dagger$ \\
\texttt{two\_safe}     & 0.500 & \textbf{0.500} & \textbf{0.500} & 0.000 & \textbf{0.500} & \textbf{0.500} & \textbf{0.500} \\
\texttt{asym\_safe}    & 0.218 & 0.293 & 0.275 & 0.332 & 0.015$^\dagger$ & \textbf{0.218} & 0.997$^\dagger$ \\
\texttt{polytope4}     & 0.161 & 0.264 & 0.246 & 0.332 & 0.001$^\dagger$ & \textbf{0.162} & 1.000$^\dagger$ \\
\texttt{kuhn}          & 0.201 & 0.005$^\dagger$ & 0.068 & \textbf{0.203} & 0.055$^\dagger$ & 0.180 & 0.000$^\dagger$ \\
\bottomrule
\end{tabular}
\end{table}

The Jensen--Shannon divergences to R-NaD corroborate the coordinate story in a coordinate-free way: they are essentially zero on the symmetric games and grow on the asymmetric ones (e.g.\ CFR\textsuperscript{+}-to-R-NaD JS of $0.009$ on \texttt{asym\_safe} and $0.016$ on \texttt{polytope4}), while MWU's divergence is an order of magnitude larger. The $L_0$ support distinguishes the collapse of MWU (support ${\approx}1.1$--$1.5$) from the well-mixed regularized and averaging solutions (support $2$--$3$).

\subsection{Geometric mapping}
\label{sec:geom}
\begin{figure}[t]
\centering
\includegraphics[width=0.95\textwidth]{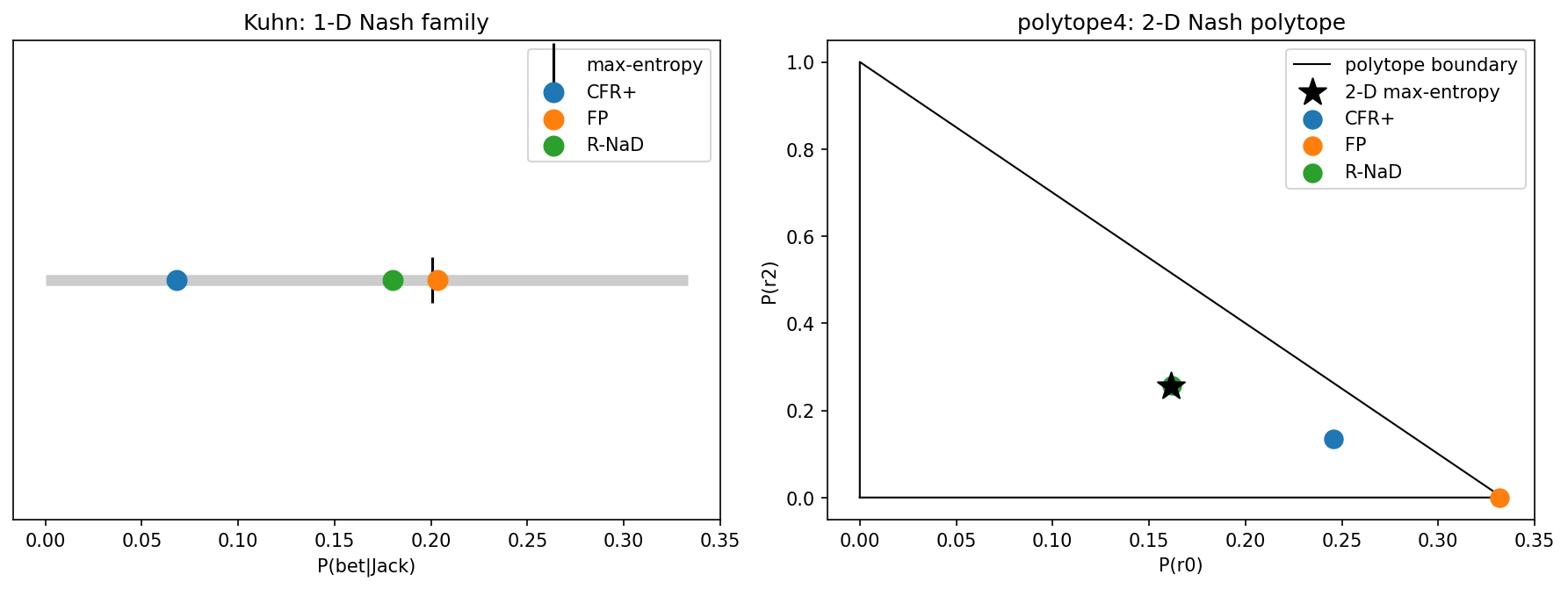}
\caption{Where each converged solver lands on the actual Nash set. Left: Kuhn's 1-D family (bluff frequency). Right: the 2-D Nash polytope of \texttt{polytope4}. R-NaD coincides with the max-entropy point; CFR\textsuperscript{+} does not.}
\label{fig:geomap}
\end{figure}

Figure~\ref{fig:geomap} draws the landing points of the converged solvers directly on the Nash set: the 1-D Kuhn family and the 2-D \texttt{polytope4} face. In both, R-NaD lands on the max-entropy point (the star/tick), and CFR\textsuperscript{+} lands away from it; in the 2-D case the separation is visible in the interior of the polytope rather than along a single axis, ruling out the possibility that ``max-entropy'' is merely a 1-D midpoint artefact.

\subsection{Bias versus convergence: a fixed magnet is not enough}
\label{sec:exp1}
\begin{figure}[t]
\centering
\includegraphics[width=0.95\textwidth]{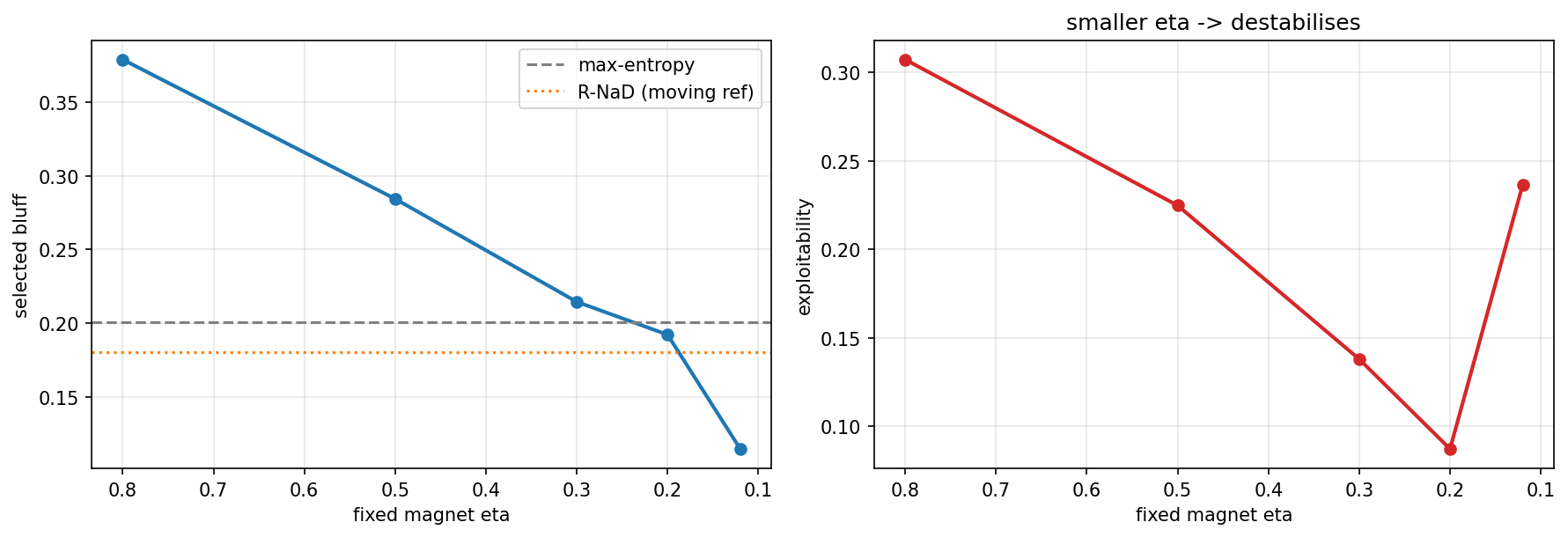}
\caption{Bias versus convergence on Kuhn. Left: as the \emph{fixed} magnet $\eta$ decreases (right to left), the selected coordinate approaches max-entropy. Right: but exploitability blows up once $\eta\lesssim0.2$. R-NaD's moving reference (dotted) avoids the trade-off, converging exactly at $99.7\%$ of maximum entropy.}
\label{fig:exp1}
\end{figure}

A natural hypothesis is that the small residual Kuhn gap (R-NaD bluff $0.180$ vs.\ max-entropy $0.201$) could be closed by annealing the magnet strength $\eta\!\to\!0$, recovering the exact maximum-entropy point. It cannot. Figure~\ref{fig:exp1} sweeps a \emph{fixed} magnet: as $\eta$ shrinks the selected coordinate does move toward max-entropy, but exploitability rises sharply once $\eta\lesssim 0.2$---the regularization that pins the dynamics is also what stabilises it, and below this threshold the unregularized dynamics enter limit cycles \citep{mertikopoulos2018,bailey2018}. The \emph{moving} reference of R-NaD is what reaches an exact Nash ($\textsc{NashConv}=0$) while still attaining $99.7\%$ of the maximum entropy. The residual coordinate gap is the flat top of the entropy landscape---near-identical entropy, slightly different coordinate---not a convergence shortfall to be optimised away. We therefore report this as a bias--stability frontier rather than a defect.

\subsection{The two-dimensional polytope}
\label{sec:exp2}
On \texttt{polytope4} the Nash set is genuinely two-dimensional, so the max-entropy member is a non-trivial interior point $(0.161,0.256)$ rather than a segment midpoint. (Throughout, the scalar selection coordinate reported for this game in Table~\ref{tab:core} is the first component $p_0$, player~0's probability on row action $r_0$; the second component $p_2$ is reported only in this section.) R-NaD lands at $(0.162,0.257)$---max-entropy to three decimals in both coordinates and with exploitability $0$---whereas CFR\textsuperscript{+} lands at $(0.246,0.135)$ with non-zero residual exploitability. This is the cleanest single confirmation that the regularized fixed point computes the I-projection \eqref{eq:iproj} of the uniform reference, since on a 2-D set the claim has two independent degrees of freedom to satisfy.

\subsection{Generalization across a random game ensemble}
\label{sec:ensemble}
The six games above are hand-built, so a natural worry is that the regularized-vs-averaging split is a property of those specific constructions rather than of the algorithms. We test this on a randomized population. We generate $N=180$ \emph{random asymmetric safe-action} games: a random $2\times2$ zero-sum core with a verified unique interior mixed equilibrium (value $V^\star$, computed in closed form), augmented with $k\in\{1,2,3\}$ safe rows equal to $[V^\star,V^\star]$. Core entries are sampled at magnitude $\mathcal{U}[0.5,3.0]$ (diagonal positive, off-diagonal negative), and a draw is retained only when its unique mixed equilibrium is interior---both equilibrium probabilities in $(0.05,0.95)$; the three values $k\in\{1,2,3\}$ contribute $60$ games each. Because the safe rows are worth $V^\star$ against any opponent, player~1's indifference is governed only by the two active rows; holding them in the equilibrium ratio keeps player~1 at its unique strategy, so each game has an analytically known, \emph{skewed} Nash face whose max-entropy member---the I-projection of the uniform reference---we compute in closed form. Every face member is Nash to machine precision ($<10^{-15}$).\footnote{R-NaD places the learning rate inside the softmax exponent \eqref{eq:mmd} and is therefore sensitive to payoff scale; on the ensemble we set $\eta=1/\mathrm{range}(M)$ per game. This is a \emph{stability} requirement, not a selection choice---regret matching is scale-invariant and needs no such adjustment, and once R-NaD reaches a Nash its selected member is unchanged. At a fixed $\eta=1$, R-NaD limit-cycles on games with large payoff range (the instability of \S\ref{sec:exp1}), failing to converge on $\sim$half the ensemble; with the rescaling it converges on all of it. CFR\textsuperscript{+} converges on $90\%$ of the games at a fixed budget.}

\begin{figure}[t]
\centering
\includegraphics[width=0.95\textwidth]{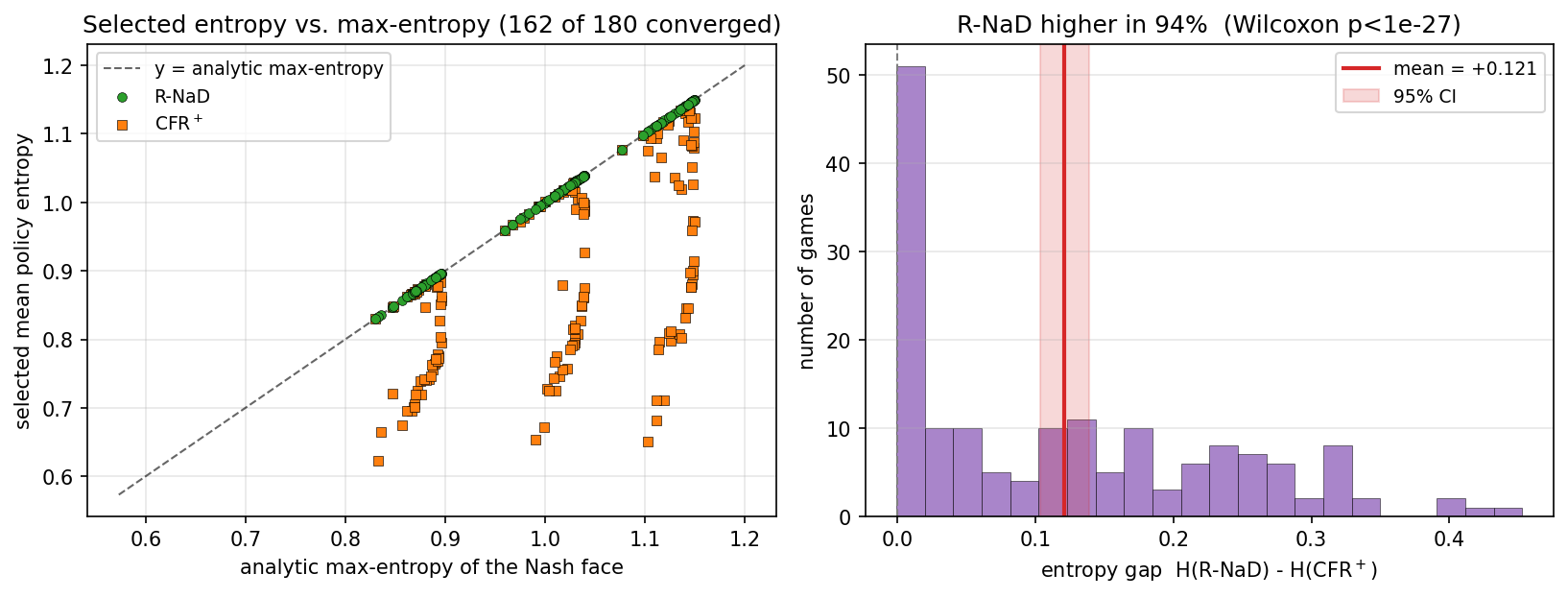}
\caption{Generalization across a $180$-game random ensemble of asymmetric safe-action games. \textbf{Left:} selected mean policy entropy versus the analytic max-entropy of each game's Nash face; R-NaD (green) lies on the diagonal (it \emph{is} the max-entropy member), CFR\textsuperscript{+} (orange) lies strictly below. The three diagonal bands correspond to $k\in\{1,2,3\}$ safe rows. \textbf{Right:} distribution of the entropy gap $H(\text{R-NaD})-H(\text{CFR\textsuperscript{+}})$ over the $162$ games where both solvers converge; mean $+0.121$, $95\%$ bootstrap CI shaded, paired Wilcoxon $p<10^{-27}$.}
\label{fig:ensemble}
\end{figure}

The split is stark and consistent (Figure~\ref{fig:ensemble}). Across the $162$ games on which both solvers converge, R-NaD lands on the analytic max-entropy member in \emph{$100\%$} of games (median coordinate error $2\times10^{-4}$, maximum $6\times10^{-4}$; selected entropy indistinguishable from the maximum), a population-scale confirmation of the I-projection characterization \eqref{eq:iproj}---with two independent degrees of freedom satisfied on the $k\ge2$ faces. CFR\textsuperscript{+}, by contrast, sits strictly below the maximum-entropy member in $94\%$ of games, with a mean entropy gap $H(\text{R-NaD})-H(\text{CFR\textsuperscript{+}})=+0.121$ ($95\%$ bootstrap CI $[+0.103,+0.139]$). Treating each game as one paired observation, a Wilcoxon signed-rank test\footnote{We use the Wilcoxon signed-rank test rather than a paired $t$-test because the per-game entropy gap is strongly right-skewed and floored near zero---R-NaD attains the max-entropy member (the maximum over the face), so the gap is essentially non-negative, with a spike of boundary ties and a long right tail---which violates the normality a $t$-test assumes. The signed-rank test is nonparametric yet, unlike the sign test, still uses the magnitude of each paired difference. The sign test (sign only) and the distribution-free bootstrap confidence interval are reported alongside and reach the same conclusion; the alternative is one-sided because the direction is predicted by the theory. The comparison is paired (same game, two solvers), which rules out unpaired tests.} rejects the null of no gap at $p<10^{-27}$ (sign test $153/161$, $p<10^{-35}$). The few ties are games whose max-entropy member lies at a face boundary, where any converged solver must coincide. The phenomenon of \S\ref{sec:core} is therefore not an artifact of the six chosen games but a property of the algorithm families on asymmetric Nash polytopes.

\subsection{Convergence bake-off}
\label{sec:bakeoff}
\begin{figure}[t]
\centering
\includegraphics[width=0.92\textwidth]{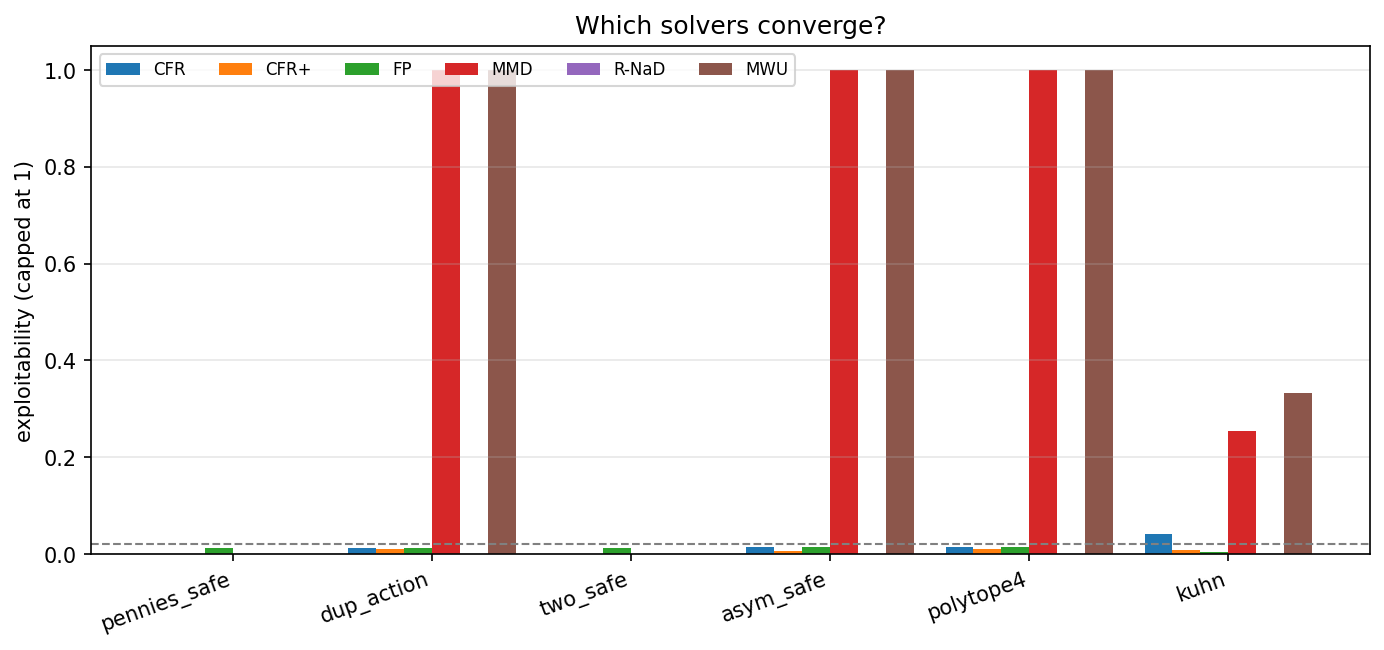}
\caption{Exploitability (capped at $1$) by solver and game; dashed line is the convergence threshold. Fixed-magnet MMD and unregularized MWU fail to converge on asymmetric games.}
\label{fig:conv}
\end{figure}

Selection is only meaningful for solvers that actually reach a Nash equilibrium. Figure~\ref{fig:conv} reports exploitability (capped at $1$) across games. Regret-averaging (CFR, CFR\textsuperscript{+}, and---once its slow $O(1/\sqrt{T})$ averaging is given enough iterations---FP) and the moving-reference regulariser R-NaD converge on the matrix games; a \emph{small fixed} magnet (MMD) and the unregularized last iterate (MWU/NeuRD) do not, exhibiting the limit-cycling characteristic of unregularized learning in zero-sum games. Vanilla CFR with uniform averaging converges slowly and does not reach the threshold on Kuhn within budget; CFR\textsuperscript{+} does. These non-convergences are themselves informative: they are the empirical justification for a moving reference, which is required to satisfy \emph{both} zero exploitability \emph{and} max-entropy selection.

\subsection{Selection is algorithmic, not seed- or budget-dependent}
\label{sec:seed}
The tabular solvers are \emph{deterministic}: they iterate on exact counterfactual values from a fixed uniform start, with no sampling, so the random seed has no effect on the dynamics. Selection is therefore seed-invariant by construction rather than by luck; re-running CFR\textsuperscript{+} and R-NaD on \texttt{asym\_safe} across four seeds confirms this at the implementation level---the within-algorithm standard deviation of the selection coordinate is $0.00000$ for both (means $0.275$ and $0.218$)---ruling out any hidden run-to-run nondeterminism. The differences in Table~\ref{tab:core} are thus a property of the update rule. Sensitivity to the \emph{initial reference} is a separate question---the seed does not set it---and is characterized in \S\ref{sec:anchor}, where biasing the reference does move the selection.

\begin{figure}[t]
\centering
\includegraphics[width=0.6\textwidth]{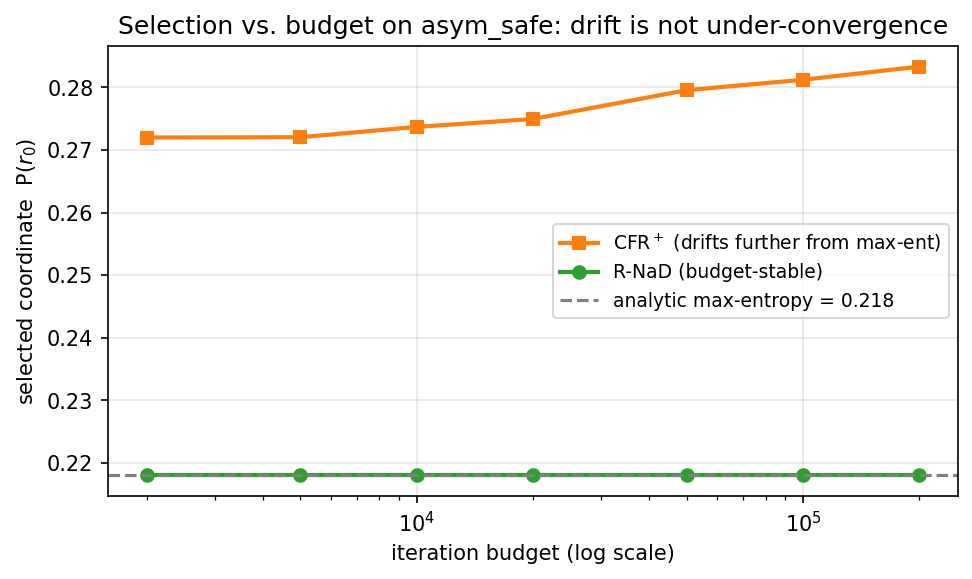}
\caption{Selection versus iteration budget on \texttt{asym\_safe} (log $x$-axis). R-NaD is pinned at the analytic max-entropy coordinate $0.218$ at every budget; CFR\textsuperscript{+} drifts \emph{away} from it as the budget grows. The family split is not under-convergence.}
\label{fig:budget}
\end{figure}

A second confound is the fixed iteration budget: perhaps the regret-averaging methods would reach max-entropy given more iterations, and their drift is merely incomplete convergence. They would not. Figure~\ref{fig:budget} sweeps the budget on \texttt{asym\_safe} over two orders of magnitude. R-NaD's coordinate is invariant at the max-entropy value $0.218$ throughout, while CFR\textsuperscript{+}'s drifts \emph{further} from max-entropy as the budget grows ($0.272\!\to\!0.283$ from $2{,}000$ to $200{,}000$ iterations), even as its exploitability continues to fall. The drift is therefore a genuine selection property of regret-averaging dynamics, not an under-resourced sweep: more compute moves CFR\textsuperscript{+} away from, not toward, the max-entropy member.

\subsection{Anchor-following: the selection depends on the reference}
\label{sec:anchor}
\begin{figure}[t]
\centering
\includegraphics[width=0.7\textwidth]{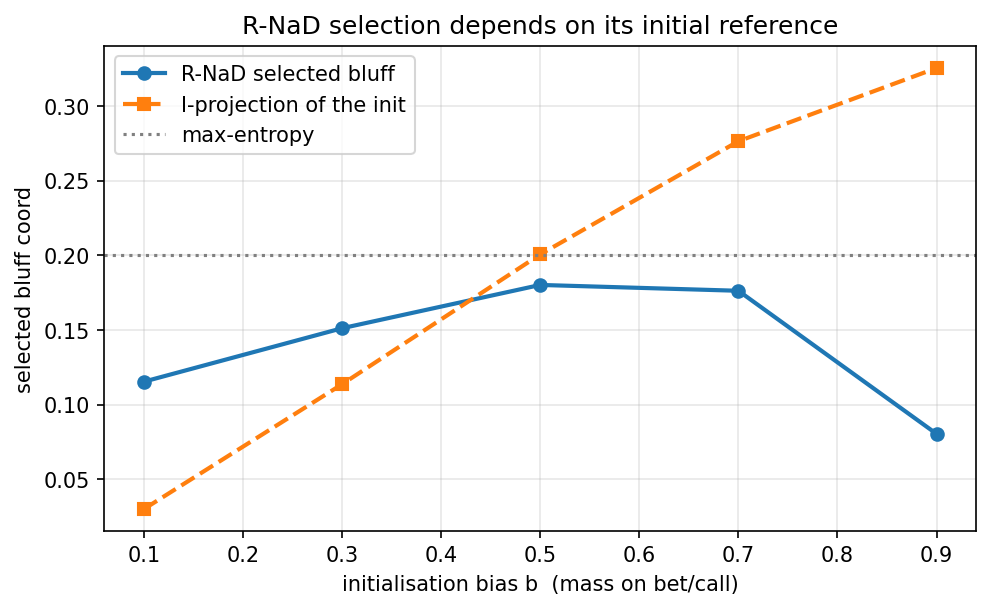}
\caption{R-NaD on Kuhn is anchor-following. As the initial reference is biased ($x$-axis), the selected equilibrium (solid) shifts over $0.08$--$0.18$ (all exact Nash), tracking the I-projection of the reference (dashed) in direction. Max-entropy (dotted) is recovered at a uniform reference, not unconditionally.}
\label{fig:init}
\end{figure}

Is R-NaD's max-entropy selection an unconditional structural attractor, or does it depend on the reference? Figure~\ref{fig:init} sweeps the \emph{initial reference} on Kuhn (decoupling it from the magnet), biasing the starting policy from $b=0.1$ to $0.9$ toward the betting action. The selected coordinate is \emph{not} invariant: it moves over the range $0.08$--$0.18$, all at exploitability $0$ (every selected profile is an exact Nash). R-NaD is therefore \emph{anchor-following}---it is regularized toward its initial reference, which it then tracks. With a \emph{uniform} initial reference (the standard default) it selects the (near-)max-entropy member; biasing the reference moves the selection, in the direction of the I-projection of the reference for moderate biases, though the precise tracking is approximate because the reference moves during training. (At extreme biases the selected coordinate is in fact non-monotone---it rises and then falls while the I-projection rises throughout, Figure~\ref{fig:init}---reflecting that the deployed limit follows the \emph{trajectory} of resets rather than a single cold projection of $\rho_0$; we therefore do not fit a linear tracking law and read the relation as directional, not exact, off uniform.) This refines the characterization: R-NaD computes the I-projection of its \emph{reference}, and max-entropy selection is the special case of a uniform reference, not an unconditional property. It also explains why the fixed-biased-magnet variant of MMD is an unreliable probe of the same mechanism: with a fixed off-uniform magnet, MMD fails to converge (\S\ref{sec:bakeoff}), confounding the selection it would otherwise reveal.

\subsection{Is the $\max(R,0)$ projection the cause of boundary drift? A controlled refutation}
\label{sec:hedge}
A common explanation for CFR-family boundary drift is that regret matching's positive-orthant projection $\max(R,0)$ structurally favours sparsity, dragging the average toward low-entropy faces. We test this directly with \emph{Hedge}, which uses the \emph{same} counterfactual regrets as CFR but replaces the hard projection with a softmax, $\sigma\propto\exp(\eta R)$, retaining strategy averaging. If the projection were the cause, removing it should move the selection toward max-entropy. It does the opposite: on \texttt{asym\_safe}, CFR selects coordinate $0.293$ (entropy $0.796$) while Hedge selects $0.334$ (entropy $0.649$); the same direction holds on \texttt{polytope4} and Kuhn. Removing the clamp drifts \emph{further} to the boundary, at \emph{lower} entropy. We conclude that the $\max(R,0)$ projection is \emph{not} the mechanism: boundary drift is a property of regret-\emph{averaging} dynamics more broadly, and its mechanism remains open. We report this negative result rather than repeat the standard intuition.

\subsection{Where downstream consequences of selection appear}
\label{sec:teeth}
\begin{figure}[t]
\centering
\includegraphics[width=0.65\textwidth]{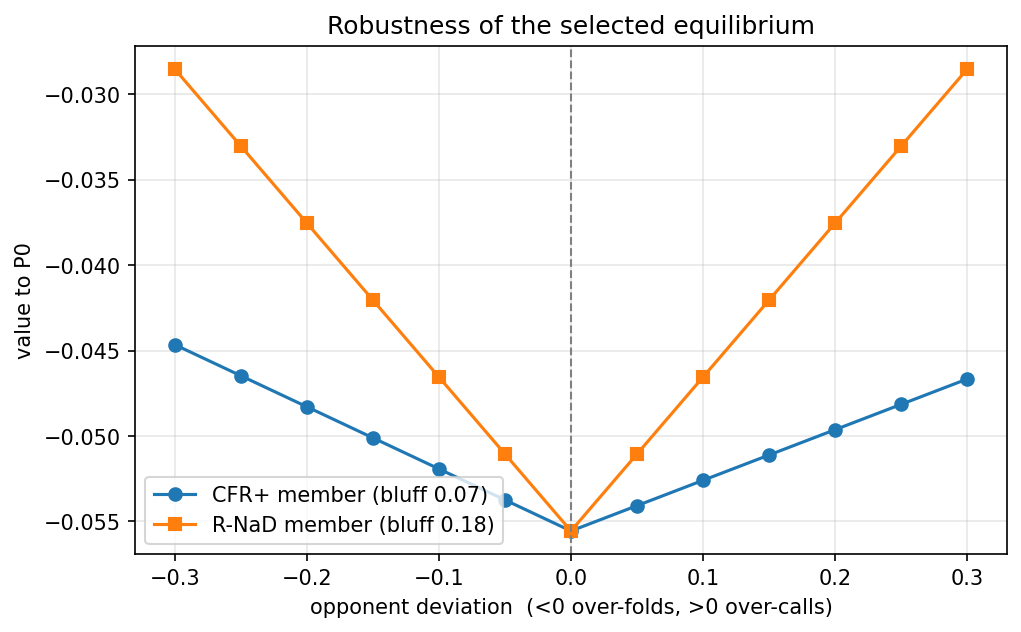}
\caption{Robustness of the selected Kuhn equilibrium. The max-entropy (R-NaD) member's value to player~0 dominates the CFR\textsuperscript{+} member's against every opponent deviation, tying only against a Nash opponent ($\delta=0$).}
\label{fig:robust}
\end{figure}

Do these differences matter? Against a best-responding opponent, no: all members share $V^\star$, so any downstream effect must be \emph{off-path}, against sub-optimal opponents. We freeze each solver's player-0 strategy and evaluate it against opponents that over-fold or over-call by a factor $\delta$. On Kuhn (Figure~\ref{fig:robust}), the R-NaD (max-entropy) member weakly dominates the CFR\textsuperscript{+} member at every deviation ($25/25$ on a fine grid) and the two tie exactly at the Nash opponent ($\delta=0$)---the max-entropy member is a strictly better hedge with no cost on-path. This dominance is, however, \emph{specific to Kuhn's sequential structure}: on the matrix games the two members differ but neither dominates---across opponent deviations CFR\textsuperscript{+} is the better hedge against $17/25$ of them on both \texttt{asym\_safe} and \texttt{polytope4}, R-NaD against the remainder. The hedging advantage of the max-entropy member is thus an extensive-form phenomenon that tracks the structure-gated ``teeth'' below, not a general property of max-entropy selection.

\begin{figure}[t]
\centering
\includegraphics[width=0.6\textwidth]{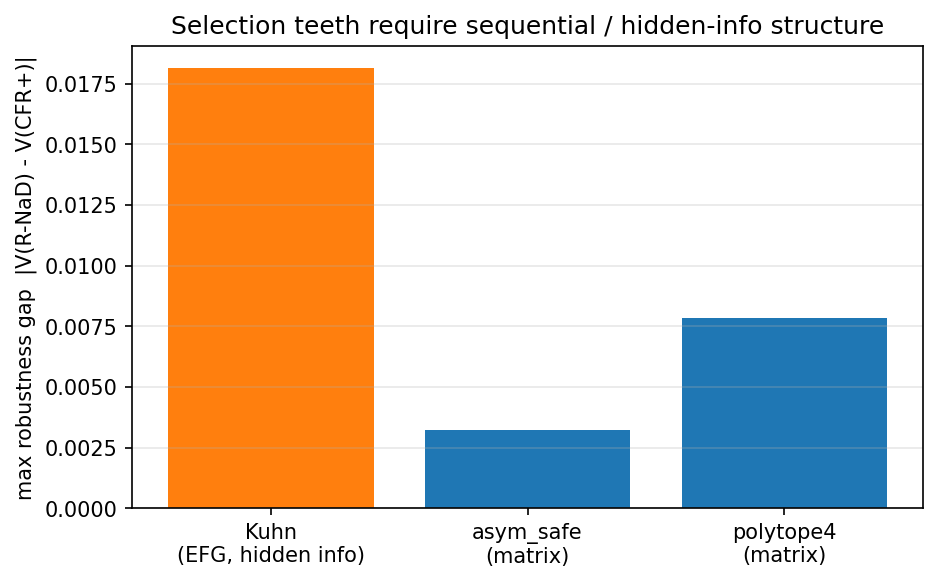}
\caption{Maximum robustness gap $|V(\text{R-NaD})-V(\text{CFR\textsuperscript{+}})|$ across opponent deviations. The gap is ${\sim}5.6\times$ larger in the hidden-information EFG (Kuhn) than in the matrix games, but bounded ($<0.02$) in absolute terms.}
\label{fig:teeth}
\end{figure}

The magnitude of this effect, however, must be stated carefully. A natural request is for a game with \emph{payoff-inequivalent} equilibria, to show selection is not cosmetic. In a 2p0s game such equilibria \emph{cannot exist}: the minimax theorem forces every Nash member to share $V^\star$ against an optimal opponent. The consequences of selection therefore live entirely in off-equilibrium exploitability, and Figure~\ref{fig:teeth} shows these scale with structure but are bounded. The robustness gap between the R-NaD and CFR\textsuperscript{+} members is $0.0181$ in Kuhn versus $0.0032$--$0.0078$ in the matrix games---about $5.6\times$ larger---yet even in Kuhn it remains under $0.02$ on a game whose value is $-1/18$. The honest reading is that sequential/hidden-information structure \emph{multiplies} the importance of selection (giving it real ``teeth'' when opponents step into unvisited subgames) while the absolute magnitude stays modest. This is a relative-scale effect, not an absolute one. Genuinely value-distinct equilibria require leaving the zero-sum setting (\S\ref{sec:future}).

\section{Discussion}
\label{sec:disc}

\paragraph{What is established.}
(i)~Equilibrium selection in 2p0s games with non-singleton Nash sets is a systematic function of the solver, not the seed (\S\ref{sec:seed}) or the iteration budget (\S\ref{sec:seed}), and the across-family differences are gated by the geometry of the Nash set---absent on symmetric sets, present on asymmetric ones (\S\ref{sec:core}). (ii)~With a uniform reference, regularized last-iterate methods select the maximum-entropy member, exactly on a 2-D polytope and to $99.7\%$ in Kuhn (\S\ref{sec:core}--\ref{sec:exp2}), and on $100\%$ of a $180$-game random ensemble of asymmetric games while CFR\textsuperscript{+} sits strictly below it ($94\%$, paired Wilcoxon $p<10^{-27}$; \S\ref{sec:ensemble}). (iii)~That selection is anchor-following, not unconditional (\S\ref{sec:anchor}). (iv)~Regret-averaging methods drift to a lower-entropy face, and this is \emph{not} caused by the $\max(R,0)$ projection (\S\ref{sec:hedge}). (v)~Downstream consequences are bounded; the max-entropy member is a strictly better hedge in Kuhn, but on the matrix games the members differ without either dominating, so the hedging advantage scales with sequential structure rather than being intrinsic to max-entropy (\S\ref{sec:teeth}).

\paragraph{What is conjectured.}
The unifying explanation we find most consistent with the data is that the regularized fixed point computes the I-projection \eqref{eq:iproj} of its reference onto the Nash polytope.

\begin{conjecture}[Regularized selection as I-projection]
\label{conj:iproj}
Let $G$ be a 2p0s game with Nash polytope $\NE(G)$ and let R-NaD (equivalently, MMD with a moving reference) be run with initial reference $\rho_0$. Then the deployed limit equals $\Iproj_{\NE}(\rho_0)=\argmin_{\sigma\in\NE(G)}\KL(\sigma\Vert\rho_0)$. In particular, with $\rho_0=\mathrm{unif}$ the limit is the maximum-entropy equilibrium $\sigma^{\mathrm{ME}}$.
\end{conjecture}

\noindent The evidence is strong and consistent (exact on every matrix game and on the 2-D polytope; $99.7\%$ in Kuhn; an exact match on $100\%$ of a $180$-game random ensemble under uniform initialization, \S\ref{sec:ensemble}; the anchor-following sweep of \S\ref{sec:anchor} matches the I-projection in direction over moderate biases), but it is evidence, not proof. The matrix-game case of the conjecture is, however, not mysterious: a 2p0s matrix game is a linear program, so the maximum-entropy / I-projection selection is the expected vanishing-regularization limit of an entropic penalty \citep{weed2018,cuturi2013}, and the moving reference is the algorithmic counterpart of the logit-QRE homotopy whose endpoint is the limiting logit equilibrium \citep{turocy2005,gemp2022}. What is genuinely open is the \emph{extensive-form} case under a moving reference---where dilated entropy and the realization weighting of behavioural strategies obstruct a direct reduction---and the precise (approximate) anchor-following relationship off uniform. We accordingly frame the contribution as identifying, against analytic ground truth and across solver families, that this regularization-induced selection coincides with the I-projection of the reference, rather than as a claim that the selection principle is itself novel. We deliberately do \emph{not} assert two mechanisms that intuition suggests: that CFR's boundary drift is caused by $\max(R,0)$ (refuted, \S\ref{sec:hedge}), and that the Kuhn-only max-entropy behaviour of fictitious play reflects a general property of realization-weighted averaging (it does not---FP drifts to the boundary on all five matrix games and matches max-entropy only on Kuhn; we treat this as a single-game anomaly).

\paragraph{Reconciling the exact uniform-init match with approximate anchor-following.}
A careful reader may object that \S\ref{sec:anchor} undercuts the conjecture: if R-NaD only \emph{approximately} tracks the I-projection of a \emph{biased} reference, why read the \emph{uniform}-reference result as exact rather than merely approximate-but-close? Three points distinguish the two regimes. First, the quality of evidence differs in kind, not degree: under a uniform reference the selection coincides with $\sigma^{\mathrm{ME}}$ to numerical precision on \emph{five} matrix games and, decisively, at the interior of the \emph{two}-dimensional polytope $(0.162,0.257)$ vs.\ $(0.161,0.256)$---a two-degree-of-freedom target that an approximate match would miss generically. The biased-init sweep, by contrast, reports a \emph{trend} (the selection moves in the I-projection's direction), which is a weaker measurement and is consistent with exact selection corrupted by a second effect rather than with inexact selection. Second, that second effect is identified and is specific to biased references: with a non-uniform initial reference, the moving-magnet schedule traverses a sequence of QREs anchored to a \emph{changing} reference, so the deployed limit reflects the \emph{trajectory} of references, not a single cold projection of $\rho_0$; under a uniform $\rho_0$ this confound is absent because the first reference already coincides with the max-entropy target's anchor and the reset sequence stays in its vicinity. Third, the one regime where even the uniform-init match is inexact---Kuhn, at $99.7\%$ of maximum entropy---has an independent, measured explanation that is \emph{not} inexact projection: the entropy functional is nearly flat near its maximum over the Kuhn family (\S\ref{sec:exp1}), so a profile at $99.7\%$ of maximum entropy can differ in coordinate while being essentially indistinguishable in the quantity the I-projection actually optimises. We therefore read the conjecture as: R-NaD computes the exact I-projection of its uniform reference up to the flatness of the entropy landscape, and the biased-init approximation reflects reference-trajectory dependence rather than a failure of the projection itself. Both halves of this reading are falsifiable---e.g.\ by a game with a sharply curved entropy maximum, where the conjecture predicts an \emph{exact} uniform-init coordinate match---and we flag the test as future work.

\paragraph{Averaging is mostly monolithic.}
An earlier reading of partial data suggested ``averaging is not monolithic'' because FP appeared to reach max-entropy. With convergence tightened, that reading does not survive: CFR and CFR\textsuperscript{+} drift to the boundary on the asymmetric games, and FP drifts to the boundary on \emph{every} matrix game---including the symmetric ones, where the other converging solvers reach max-entropy. FP's max-entropy match is confined to Kuhn, and we treat it as an unexplained single-game anomaly rather than evidence that realization-weighted averaging is qualitatively different.

\section{Limitations}
\label{sec:limits}
Our testbed is tabular and small, chosen so that ground truth is exact; we do not study function-approximation or sampling effects, which could interact with selection. Conjecture~\ref{conj:iproj} is unproven. The setting is strictly zero-sum, so the strongest possible form of ``selection matters''---different \emph{values} under different selections---is out of scope by the minimax theorem. The downstream effect, while structurally robust, is small in absolute terms, and the max-entropy member's robustness \emph{advantage} is confined to the extensive-form game (Kuhn); on matrix games neither member dominates. Our random ensemble (\S\ref{sec:ensemble}) broadens the evidence over \emph{matrix} games only---Kuhn remains the sole extensive-form instance, so claims that scale \emph{with} sequential structure rest on a single EFG; a randomized EFG family is the natural next step. R-NaD's last-iterate convergence (though not its selected member) is sensitive to payoff scale through the learning rate, requiring a per-game $\eta\propto1/\mathrm{range}(M)$ on the ensemble; regret matching needs no such tuning. Finally, the anchor-following result means our max-entropy claims are conditional on uniform initialization; deployments that warm-start from a non-uniform reference will select a different member.

\paragraph{Scope and resources.} This is an independent research project carried out on a single personal computer (commodity CPU, no GPU or compute cluster). Several of the scoping choices above are therefore as much practical as principled: the tabular, exactly solvable testbed was chosen both because it admits analytic ground truth \emph{and} because it runs end-to-end on a laptop in minutes; the absence of function-approximation and sampling (e.g.\ Monte-Carlo CFR or deep) experiments, and the restriction of the random ensemble to matrix games with a single extensive-form instance, reflect this compute budget rather than a belief that the phenomena vanish at scale. We have tried to make the constraint a virtue---every number and figure is reproducible from exact counterfactual values on modest hardware---and we flag scaling these experiments to function-approximation regimes and larger extensive-form games as the natural, resource-permitting extension.

\section{Related work}
\label{sec:related}
Regret minimization and its averaging guarantee underlie CFR \citep{zinkevich2007} and CFR\textsuperscript{+} \citep{tammelin2014,bowling2015}, which deploy time-averaged strategies. Fictitious play \citep{brown1951} converges in 2p0s games \citep{robinson1951}. A separate line regularizes the learning dynamics toward a reference policy to obtain last-iterate convergence: Neural Replicator Dynamics \citep{hennes2020}, the regularization-for-convergence analysis of \citet{perolat2021}, R-NaD and DeepNash \citep{perolat2022}, and Magnetic Mirror Descent \citep{sokota2023}, which unifies regularized RL, quantal-response equilibria \citep{mckelvey1995}, and 2p0s solving. The instability of \emph{unregularized} dynamics---limit cycles and Poincar\'e recurrence---is well documented \citep{mertikopoulos2018,bailey2018}, and is the phenomenon our \S\ref{sec:exp1} and \S\ref{sec:bakeoff} reproduce. Equilibrium selection as a refinement question dates to \citet{harsanyi1988}; our selection principle is informational: the I-projection / maximum-entropy member \citep{jaynes1957,csiszar1975}. That entropic regularization selects the maximum-entropy member of an optimal set is established outside game solving---in linear programming the vanishing-penalty limit is the maximum-entropy point of the optimal face \citep{weed2018}, and entropic optimal transport selects the KL projection of a reference onto the feasible polytope \citep{cuturi2013}---so Conjecture~\ref{conj:iproj} is best read as the 2p0s instantiation of this principle rather than a new principle. On the game-theoretic side, the moving reference traces the logit-QRE homotopy whose zero-temperature endpoint is the \emph{limiting logit equilibrium} \citep{mckelvey1995,turocy2005}, the selection target of homotopy solvers such as ADIDAS \citep{gemp2022}; that homotopy literature is generically concerned with games having a \emph{unique} endpoint, whereas our degenerate (polytope) testbed is exactly the non-generic case it sets aside, and which we characterize as the I-projection. That different learning rules can select different equilibria is itself a known theme \citep{harsanyi1988}; relatedly, no-regret/FTRL last iterates are known to be unstable at non-strict (mixed) Nash equilibria \citep{flokas2020}, consistent with the boundary drift we observe for the regret-averaging solvers. Against this backdrop, the new contributions here are: the systematic, ground-truth-verified comparison of \emph{which} Nash-polytope member each solver family selects---on hand-built games with analytically known Nash sets, including a 2-D polytope and Kuhn, and on a randomized $180$-game ensemble---together with the two negative results, on the $\max(R,0)$ mechanism and on initialization-independence. The I-projection characterization itself we contribute as the game-solving instance of the regularization-selection principle above, supported at population scale and stated as a falsifiable conjecture.

\section{Future work}
\label{sec:future}
Four directions follow directly. (1)~\textbf{Prove Conjecture~\ref{conj:iproj}}---characterise the moving-reference limit as the I-projection of the reference onto the Nash polytope; the empirics here are its backbone. (2)~\textbf{Explain regret-averaging boundary drift}, whose obvious explanation ($\max(R,0)$) we have refuted. (3)~\textbf{General-sum extension}: genuinely payoff-inequivalent equilibria---different values against an optimal opponent---exist only outside zero-sum, where minimax no longer applies and R-NaD/CFR lose their guarantees; selection there is the natural frontier, but a distinct setting. (4)~\textbf{Amplified-stakes games}: a custom extensive-form game can enlarge the off-path swing, but should be presented only as a labelled ceiling illustration alongside an untuned benchmark such as Kuhn, not as a headline---a tuned game trades away the credibility an untuned benchmark provides.

\section{Conclusion}
Solvers that are guaranteed to find \emph{a} Nash equilibrium do not find \emph{the same} one. On asymmetric Nash polytopes, regularized last-iterate methods select the maximum-entropy member (the I-projection of a uniform reference)---which we confirm on $100\%$ of a $180$-game random ensemble---regret-averaging methods drift to a lower-entropy face, and the choice is a property of the algorithm rather than the seed or the budget. In the sequential, hidden-information setting (Kuhn) the maximum-entropy member is a strictly better hedge against flawed opponents; on matrix games the members differ without either dominating, so this robustness advantage tracks sequential structure rather than max-entropy per se, and is bounded in magnitude throughout. We have been explicit about the line between demonstrated phenomena and conjectured mechanisms, and have reported two negative results---on the $\max(R,0)$ projection and on initialization-independence---that correct common intuitions. The central characterization is offered as a precise, falsifiable, strongly data-supported conjecture.

\paragraph{Reproducibility.}
All games, solvers, metrics, and the random-ensemble generator are implemented in a single self-contained tabular engine (\texttt{efg.py}) and an executable notebook that regenerates every number and figure in this paper from exact counterfactual values, with no sampling or function approximation.

\appendix
\section{Full metric table}
\label{app:metrics}
Table~\ref{tab:full} reports, for every (game, solver), the selection coordinate, mean policy entropy, and exploitability. Daggers in Table~\ref{tab:core} correspond to the exploitability values here above $0.02$.

\begin{table}[h]
\centering
\caption{Full metrics: coordinate / entropy / exploitability.}
\label{tab:full}
\scriptsize
\setlength{\tabcolsep}{3.5pt}
\resizebox{\textwidth}{!}{%
\begin{tabular}{llcccccc}
\toprule
Game & metric & CFR & CFR\textsuperscript{+} & FP & MMD & R-NaD & MWU \\
\midrule
\multirow{1}{*}{\texttt{pennies\_safe}}
 & coord/$H$/expl & 0.333/0.90/0.00 & 0.333/0.90/0.00 & 0.000/0.69/0.01 & 0.333/0.90/0.00 & 0.333/0.90/0.00 & 0.333/0.90/0.00 \\
\texttt{dup\_action}
 & coord/$H$/expl & 0.248/0.87/0.01 & 0.250/0.87/0.01 & 0.504/0.69/0.01 & 0.008/0.24/1.72 & 0.250/0.87/0.00 & 0.000/0.00/2.00 \\
\texttt{two\_safe}
 & coord/$H$/expl & 0.500/1.04/0.00 & 0.500/1.04/0.00 & 0.000/0.69/0.01 & 0.500/1.04/0.00 & 0.500/1.04/0.00 & 0.500/1.04/0.00 \\
\texttt{asym\_safe}
 & coord/$H$/expl & 0.293/0.80/0.01 & 0.275/0.81/0.01 & 0.332/0.64/0.01 & 0.015/0.39/3.47 & 0.218/0.85/0.00 & 0.997/0.01/1.99 \\
\texttt{polytope4}
 & coord/$H$/expl & 0.264/0.91/0.02 & 0.246/0.94/0.01 & 0.332/0.64/0.01 & 0.001/0.52/2.99 & 0.162/1.00/0.00 & 1.000/0.00/2.00 \\
\texttt{kuhn}
 & coord/$H$/expl & 0.005/0.18/0.04 & 0.068/0.23/0.01 & 0.203/0.26/0.00 & 0.055/0.25/0.25 & 0.180/0.26/0.00 & 0.000/0.00/0.33 \\
\bottomrule
\end{tabular}}
\end{table}

\end{document}